# Pressure-induced ideal Weyl semimetal state in the layered antiferromagnet EuCd$_2$As$_2$

Zhenhai Yu[1,2,†], Xuejiao Chen[3,†], Wei Xia[1,4], Ningning Wang[5], Xiaodong Lv[3], Xiaolei Liu[1], Hao Su[1], Zhongyang Li[1], Desheng Wu[5], Wei Wu[5], Ziyi Liu[5,6], Jinggeng Zhao[6], Mingtao Li[2], Shujia Li[2], Xin Li[2], Zhaohui Dong[7], Chunyin Zhou[7], Lili Zhang[7], Xia Wang[8], Na Yu[8], Zhiqiang Zou[8], Jianlin Luo[5], Jinguang Cheng[5], Lin Wang[2,9,*], Zhicheng Zhong[3,*], and Yanfeng Guo[1,4,*]

[1]School of Physical Science and Technology, ShanghaiTech University, Shanghai, 201210, China

[2]Center for High Pressure Science and Technology Advanced Research, Shanghai, 201203, China

[3]CAS Key Laboratory of Magnetic Materials and Devices & Zhejiang Province Key Laboratory of Magnetic Materials and Application Technology, Ningbo Institute of Materials Technology and Engineering, Chinese Academy of Sciences, Ningbo 315201, China

[4]ShanghaiTech Laboratory for Topological Physics, ShanghaiTech University, Shanghai 201210, China

[5]Beijing National Lab for Condensed Matter Physics, Institute of Physics, Chinese Academy of Sciences, Beijing 100190, China

[6]School of Physics, Harbin Institute of Technology, Harbin 150080, China

[7]Shanghai Synchrotron Radiation Facility, Shanghai Advanced Research Institute, Chinese Academy of Sciences, Shanghai 201204, China

[8]Analytical Instrumentation Center, School of Physical Science and Technology, ShanghaiTech University, Shanghai 201210, China

[9]Center for High Pressure Science (CHiPS), State Key Laboratory of Metastable Materials Science and Technology, Yanshan University, Qinhuangdao 066004, China

**The rich nontrivial topological phases rooted in the interplay between magnetism and topology in the layered antiferromagnet EuCd$_2$As$_2$ have captured vast attention, especially the ideal Weyl semimetal state realized in the spin-polarized ferromagnetic (FM) structure driven by a moderate external magnetic field. In this work, combining high-pressure magnetotransport measurements, structure chracterizations and first principles calculations, we find that application of pressure can also realize the ideal Weyl state in EuCd$_2$As$_2$ through driving the in-plane antiferromagnetic state across an intermediate in-plane FM state then into the out-of-plane FM state. Our high-pressure angle dispersive X-ray diffraction and X-ray absorption near-edge spectroscopy measurements excluded structure transition and/or change of Eu$^{2+}$ valence state as the sources for the magnetic phase transitions. Alternatively, the apparently reduced axial ratio (*c*/*a*) and compressed Eu-layer space distance should play important roles. Our result provides an alternative way to realize the ideal Weyl semimetal state in EuCd$_2$As$_2$ and would be instructive for the exploration of exotic topological properties in such layered magnetic topological phase with strongly competing magnetic exchanges by using high pressure.**

†The authors contributed equally to this work.
*Corresponding authors:

wanglin@ysu.edu.cn (LW),

zhong@nimte.ac.cn (ZCZ),

guoyf@shanghaitech.edu.cn (YFG)

Magnetic topological phases have arrested rapid growing research interest in recent years, because the interplay between magnetism and nontrivial topological band structure can produce extraordinary topological states such as the quantum anomalous Hall (QAH) insulator [1-3], axion insulator [4-6] and magnetic Weyl semimetal (WSM) [7-12]. Generally, in a Dirac semimetal (DSM), the gapless crossing of two doubly degenerate bands forms the Dirac point (DP), which is protected by the combination of both inversion ($P$) and time-reversal ($\mathcal{T}$) symmetries [13-16]. Once $P$ or/and $\mathcal{T}$ can be broken, the twofold band degeneracy will be lifted and the crossings of nondegenerate bands then lead to Weyl points (WPs) that always appear in pairs with inverse spin chirality [17-22]. Up to now, the hydrogen atom of a WSM, i.e. a natural ideal WSM with only one single pair of WPs exactly located at or very close to the Fermi level $E_F$ in an energy window free from other overlapping bands, is yet absent. Such ideal WSMs take the advantage of exploring the predicted exceptional quantum electronic transport properties which are solely arisen from the WPs. Though the present most WSMs are realized through breaking $P$ [7-22], the $\mathcal{T}$ breaking strategy either by external magnetic field or spontaneously through correlation effect provides explicit routes toward the realization of ideal WSM. Moreover, compared with the $P$ broken WSMs, $\mathcal{T}$ broken ones would find more opportunities for use in spintronics [23-27]. In a magnetic topological phase, the coupling between the spin and charge degrees of freedom offers opportunities to control different topological phases by magnetism, which enables the creation of ideal Weyl state through precisely manipulating the spin structure, resembling the cases in the van der Waals antiferromagnetic (AFM) topological insulators (TIs) $MnBi_2Te_4/(Bi_2Te_3)_n$ [28-30] and $MnSb_2Te_4/(Sb_2Te_3)_n$ ($n = 1, 2$) [11, 12]. These materials share similarities of strong spin-orbit coupling (SOC), low structural symmetry and long-range magnetic order, in which the spin rotation or polarization by external magnetic field can significantly alter the electronic band structure by the energy of even orders of magnitude larger than the traditional Zeeman splitting [31].

The layered triangular-lattice antiferromagnets $EuCd_2Pn_2$ (Pn = As, Sb) have captured considerable interest in recent years due to the WSM state in the spin-polarized structure [9, 10, 32-36]. $EuCd_2As_2$ is an itinerant magnet with conduction electrons from the Cd and As orbitals, and the magnetism originates from the local Eu 4*f* moments which form a long-range AFM order at $T_N \sim 9.5$ K with an A-type structure, i.e., ferromagnetic (FM) *ab* planes with in-plane lying spins stacking antiferromagnetically along the *c* axis [32]. Such A-type AFM_in structure breaks the $C_3$ symmetry and consequently gaps the DP [9, 10]. Compared with the large gap of ~ 200 meV in $MnBi_2Te_4$, the small gap of only about tens meV in $EuCd_2As_2$ requires smaller energy to be closed, thus enabling the realization of FM WSM state with polarizing the spins along a specific direction, for example, the *c* axis, by a moderate external magnetic field. Interestingly, an ideal WSM state in the *c* axis direction spin-polarized structure was suggested with a single pair of WPs close to G point along the G-A direction of the Brillouin zone, in a small window of energy free of other bands, thus providing a rare clean platform for the study of Weyl physics [10].

Here, we report the pressure-induced ideal Weyl state in $EuCd_2As_2$. The pressure drives a sequential magnetic phase transitions from the AFM_in state across an FM state with all spins lying in-plane (FM_in) and finally into the FM state with all spins pointing to the out-of-plane direction (FM_out), i.e. the *c* axis direction. Our first principles calculations and Hall effect measurements suggest an ideal Weyl state in the FM_out structure.

The details for crystal growth, magnetotransport, high-pressure transport, high pressure synchrotron angle dispersive X-ray diffraction and X-ray absorption measurements methods as well as first principles calculations of the bulk $EuCd_2As_2$ are presented in the Supplementary Information (SI), which includes references [37-51].

The Single crystal growth and basic physical characterizations of $EuCd_2As_2$ were shown in panels (a)- (f) of Fig. S1. The temperature dependent resistance *R*(*T*) of $EuCd_2As_2$ crystal measured at various pressures ranging from 0 to 24 GPa is presented in Figs. 1(a)-1(b). The *R*(*T*) at ambient pressure displays a sharp peak at ~

9.5 K signifying the AFM order. The measurements at pressure lower than 2.5 GPa was recently reported [52], which shows that the AFM order could be slightly enhanced with increasing the pressure to 1.30 GPa, while the continuous increase of pressure larger than 1.3 GPa, on the contrary, can suppress the peak to low temperature until the pressure reaches ~ 2 GPa, which then rapidly pushes the peak to high temperature again. The behavior was ascribed to a spin transition from the AFM_in to FM_in configuration around 2 GPa. Our magnetic susceptibility and isothermal magnetizations measured below 0.9 GPa presented in Figs. 1(c)-1(d) support that the AFM order could be initially enhanced by pressure. We mainly focused on the characterizations on EuCd$_2$As$_2$ under pressure larger than 2.5 GPa that had not been measured yet. Seen in Figs. 1(a)-1(b), with increasing pressure larger than 2.5 GPa, the peak signifying the FM_in order gradually shifts to high temperature, which becomes more and more broad and is eventually smeared out at high pressure larger than 19.7 GPa. The $R(T)$ then displays a semi-metallic conducting behavior over the measured temperature range.

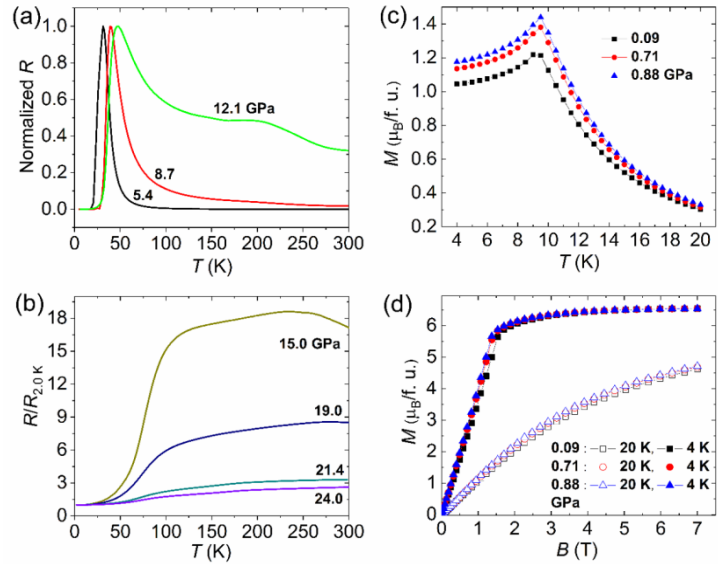

**Fig. 1.** (a)-(b) Temperature dependent resistance of EuCd$_2$As$_2$ under various pressures. (c) Temperature dependent magnetizations under pressures lower than 0.9 GPa. (d) Isothermal magnetizations at 4 K and 20 K under pressures lower than 0.9 GPa.

We also noticed that EuCd$_2$As$_2$ exhibits negative magnetoresistance (*n*-MR) under pressures in the range of 20.8 - 31.1 GPa and at 2 K, seen in Fig. 2(a). In addition, a remarkable butterfly-shaped MR was observed with increasing the pressure. To achieve more insights into the extraordinary behavior of MR under high pressure, the temperature dependence of MR at 31.1 GPa was measured and is shown in Fig. 2(b). It is interesting to note that the butterfly-shaped hysteresis loop in the MR curve was suppressed as the temperature increases. The butterfly-shaped MR could persist even up to 50 K as enlarged by the inset of Fig. 2(b). So far as we know, the observation of butterfly-shaped MR in EuCd$_2$As$_2$ has not been reported yet, which while was ever reported in other ferromagnets such as Fe$_3$O$_4$ [53] and Fe$_5$GeTe$_2$ [54]. Furthermore, a quasi-linear MR versus magnetic field was observed at 100 K. Unfortunately, since it is hard to perform crystallographic direction dependent *n*-MR measurements under very high pressure, we were unable to carry out further characterizations and analysis to achieve in-depth insights into the underlying physics.

The Hall resistance of EuCd$_2$As$_2$ at 2 K and under various pressures is shown in Fig. 2(c). Anomalous Hall effect (AHE) is displayed in the measured pressure range. Interestingly, a remarkable hysteresis loop in the Hall resistance under low magnetic field in the range of ±1.0 T was observed with increasing the pressure. Furthermore, the area of the hysteresis loop was apparently expanded with increasing the pressure. Fig. 2(d) shows the temperature dependence of Hall resistance under 31.1 GPa. The hysteresis loops were only observed in the temperature range of 2 - 10 K and then were suppressed as the temperature increases. Since it is difficult to determine the magnetization of the sample under high pressure, we can only show the overall variation tendency of the Hall resistance. We argue that the variation of the Hall resistance of EuCd$_2$As$_2$ under high pressure is correlated with the pressure-induced magnetic phase transition. High pressure neutron diffraction experiment, which could be only performed at special facilities, would be very helpful to offer more insights. Alternatively, we investigated the magnetic phase transition of EuCd$_2$As$_2$ under high pressure by employing the first principles calculations. The simulated magnetic phase transition in the sequential order of AFM_in→ FM_in → FM_out under high pressure was shown in Fig. 2(e), in which the magnetic moment of $Eu^{2+}$ is denoted with black arrows.

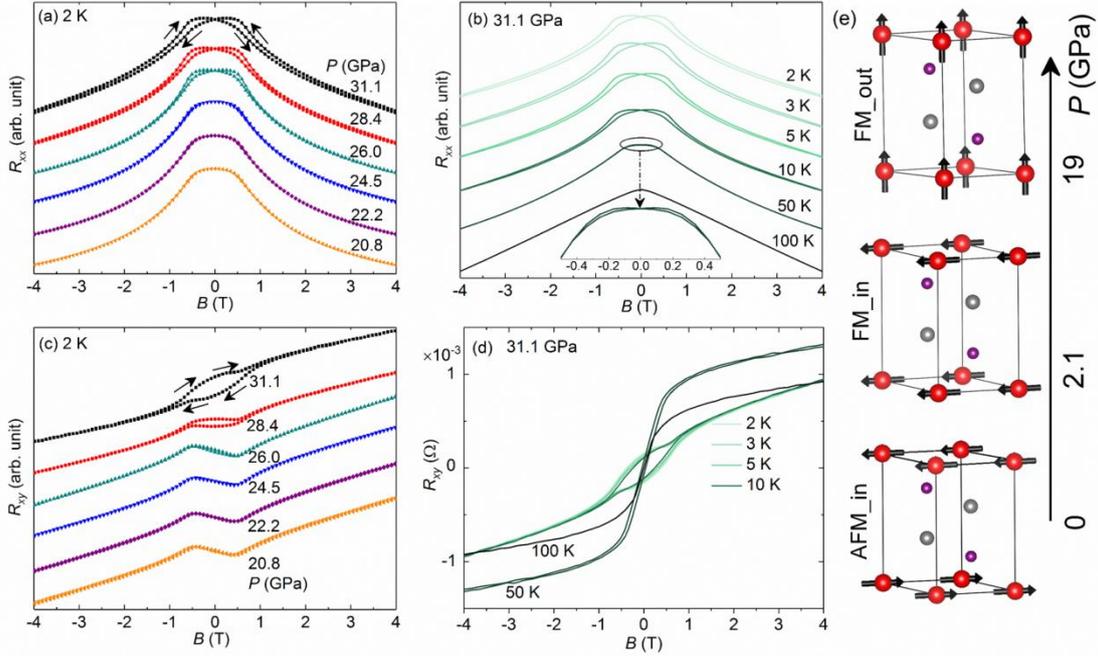

**Fig. 2.** (a) Pressure dependent MR of $EuCd_2As_2$ at 2 K. The arrows mark the direction of magnetic field variation. (b) MR under 31.1 GPa as a function of temperature. The inset shows the magnetic hysteresis within the magnetic field range of ±0.5 T at 50 K. (c) Pressure dependent Hall resistance of $EuCd_2As_2$ at 2 K. (d) Temperature dependent Hall resistance under 31.1 GPa. (e) The simulated magnetic phase transition in the sequential order of AFM_in → FM_in → FM_out under high pressure, in which the the magnetic moment of $Eu^{2+}$ is denoted with black arrows.

To examine the possibility of pressure-induced structural phase transition that is responsible for the variations of MR, Hall resistance and the magnetic phase transition, the structure evolution of $EuCd_2As_2$ under pressure were measured, with more details in the SI. At ambient pressure, the Eu, Cd and As atoms in $EuCd_2As_2$ are located at 1$a$ (0, 0, 0), 2$d$ (1/3, 2/3, 0.633), and 2$d$ (1/3, 2/3, 0.247) Wyckoff positions, respectively. The $CdAs_4$ tetrahedra are slightly elongated with Cd-As distances ranging from 2.72 to 2.84 Å. The XRD patterns below 27.1 GPa could be well indexed on the basis of the $CaAl_2Si_2$-type structure ($P$-3$m$1), seen in Fig. S2(a). The typical Rietveld refinement results of the AD-XRD patterns of $EuCd_2As_2$ collected at 0.7 and 23.8 GP are presented in Figs. S3(a) and 3(b). When the pressure is larger than 35.5 GPa, new diffraction peaks signifying the emergence of new phases appeared and the intensity of these peaks increase with further increasing the pressure, seen in Fig. S2(b). However, the major phase of the sample is still the parent $P$-$3m$1 even up to 50.0 GPa. Unfortunately, it suffers a big difficulty for refining the new peaks due to the mixture with the $P$-$3m$1 phase and the broad peaks. Interestingly, the parent phase could be recovered when the pressure was released to ambient pressure, seen in Fig. S2(b).

According to the molecular field theory [55], the magnetic ordering temperature ($T_f$) of a magnet is closely related with the effective exchange integral ($J$) between the magnetic ions. Provided that only the spin moment contributes to the magnetism within the nearest-neighbor interaction approximation, the $T_f$ could be expressed as $k_B T_f = 2zS(S+1)J/3$, where $k_B$ is Boltzmann constant, $z$ is nearest atoms number and $S$ is spin quantum number, which shows that $T_f$ is proportional to $J$. When the atomic distance is changed under certain conditions such as the application of external high pressure, $J$ and hence the $T_f$ will accordingly be varied. To trace the pressure-induced magnetic phase transitions in EuCd$_2$As$_2$, the relative compressibility ($a/a_0$ and $c/c_0$) and lattice parameters ($a$, $c$ and axial ratio $c/a$) are plotted in Figs. 3(a) and 3(b), respectively. The $P$-$3m$1 phase is a layered structure with weak chemical bonding along the $c$ axis, and the $c$ axis is therefore more compressible than the $a$ axis, as shown in Fig. 3(a). The calculated lattice parameters summarized in Fig. 3(a) agree well with the experimental ones. The pressure dependence of axial ratio $c/a$ is shown by the inset of Fig. 3(b). The general tendency of $c/a$ versus pressure exhibits a decrease with increasing pressure, which favors the competition between the AFM_in and FM_in magnetic phases. However, it is clearly visible that there is an inflection point around 2.1 GPa, which apparently signifies the magnetic phase transition from the AFM_in into the FM_in structure since the magnetic state is closely related with the evolution of crystal structure.

The $P$-$3m$1 crystal structure of EuCd$_2$As$_2$ could be visualized as a stacking of EuAs$_6$ octahedra and CdAs$_4$ tetrahedra along the $c$ axis, as illustrated in Fig. 3(c). The EuAs$_6$ octahedra are connected with each other wih edge-sharing rather than apex-sharing. Fig. 3(d) shows the pressure dependence of both Eu-As bond length and ∠As-Eu-As bond angle, which are labeled by their values. The pressure-induced EuAs$_6$ octahedron distortion between 0 and 27.1 GPa is observed, manifested by the change of ∠As-Eu-As from 90.07° to 92.49° and from 89.93° to 87.51°, respectively, as marked in Fig. 3(d). The Eu-As bond length is compressed about 8.58% as the pressure increases from 0 to 27.1 GPa. The distance between two Eu interlayers is shrunk from 7.333 Å to 6.527 Å, which is shortened by approximately 11.0% and consequently gives rise to the variation of magnetic interaction, as illustrated in Fig. 3(e). As shown in Fig. 3(a), the relative compressibility of $c/c_0$ is larger than $a/a_0$ and their difference becomes more remarkable with increasing pressure. Such crystallographic environment favors the formation of FM_out state rather than FM_in state in EuCd$_2$As$_2$ when the pressure is larger than 20 GPa.

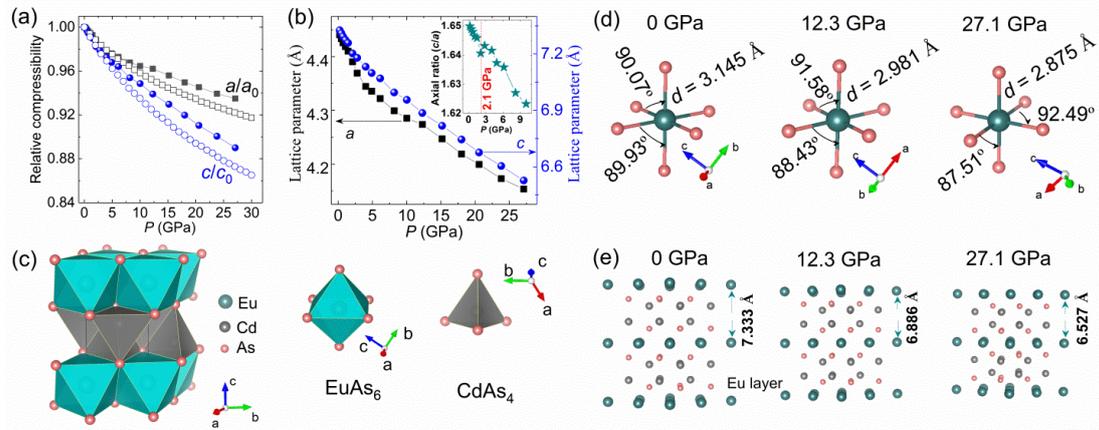

**Fig. 3.** (a) Pressure dependence of the relative compressibility $a/a_0$ and $c/c_0$ of EuCd$_2$As$_2$. The solid and open symbols correspond to experimental and calculated values, respectively. (b) Pressure dependence of lattice parameter *a* and *c*. The inset shows the axial ratio *c/a* as a function of pressure. A remarkable inflection point is observed at 2.1 GPa. (c) The crystal structure of *P*-3*m*1 phase of EuCd$_2$As$_2$. (d) The pressure dependence of Eu-As bond length and ∠As-Eu-As bond angle, from which the pressure-induced EuAs$_6$ octahedron distortion is observed. (e) The crystal structure of EuCd$_2$As$_2$ could be viewed as stacking of the Eu and CdAs layers along the *c*

axis. The distance between two Eu interlayers is shrunk as the pressure increases, which gives rise to the variation of magnetic interaction.

The possible variation of $Eu^{2+}$ ($S = 7/2$, ~ 7.0 $\mu_B$) in $EuCd_2As_2$ under pressure was also examined by measuring the high pressure XANES spectra, seen by more details in the SI. The measured Eu $L_1$–edge XANES spectra of $EuCd_2As_2$ under various pressure are presented in Fig. S4. For a comparison, the Eu $L_1$–edge XANES spectra of EuS and $Eu_2O_3$ are also plotted together. The XANES spectra for $EuCd_2As_2$, EuS, and $Eu_2O_3$ clearly demonstrate that the $L_1$–edge energy shifts depending on the oxidation state of Eu ions. As shown in Fig. S4, the linewidth of Eu $L_1$–edge becomes broader upon the increase of pressure, which might be resulted from the enhanced Eu(4$f$)-As(4$p$) hybridization or/and the intermediate valence states of Eu ions under pressure. It can be seen from Fig. S4 that the spectra profile of $EuCd_2As_2$ exhibits negligible difference from that of EuS, indicating the $Eu^{2+}$ valence state at ambient pressure. The XANES spectrum for Eu $L_1$–edge consists of two peaks, which are labeled by PE and WL, respectively. The Eu $L_1$–edge is robust against the applied pressure, indicating the stable $Eu^{2+}$ valence in $EuCd_2As_2$ under pressure. It is notable that the pre-edge peak labeled as PE remains nearly unchanged while the WL peak slightly shifts to higher photon energy with the increase of pressure. By comparing with the XANES spectrum of $Eu_2O_3$, the valence of Eu ions in $EuCd_2As_2$ at 30.6 GPa is apparently not fully trivalent but is intermediate, which is ~ +2.1. The peak positions of $EuCd_2As_2$ during compression and decompression to ambient pressure (marked as "D" in Fig. S5) are almost the same, indicating that the pressure-induced transition of +2 to intermediate valence at high pressure is reversible.

First principle calculations were performed to investigate the magnetic phase transition and the electronic band structure topology in $EuCd_2As_2$ under high pressure, see details in the SI. Figs. 4(a)-4(j) present the calculated band structures for three types of magnetic ground states of $EuCd_2As_2$. Figs. 4(a) and 4(b) are the band structures of $EuCd_2As_2$ at ambient pressure and 1.06 GPa with the AFM_in structure. At ambient pressure, a small gap about 0.025 eV at the G point of the Brillouin zone could be recognized, seen by the enlarged view in Fig. 4(e), which slightly decreases to 0.022 eV at 1.06 GPa, seen by the enlarged view in Fig. 4(f). The gap is still visible even in the FM_in structure, seen by the enlarged views in Figs. 4(g) and 4(h) taken from the band structures in Figs. 4(c) and 4(d) calculated at 2.78 and 10.02 GPa, respectively. However, with further increase of pressure, the gap is gradually closed with both electron- and hole-type bands crossing the $E_F$, seen in Figs. 4(i) and 4(j) with the pressure of 18.01 GPa and 20.57 GPa, respectively. When the pressure is lower than 18.01 GPa, the total energy calculations suggest that the FM_in state is the most stable one as compared with the other two magnetic configurations. It is clear that the band structures of the FM_in state show clear band inversion at the G point, seen in both Figs. 4(c) and 4(d). Fig. 4(k) shows the calculated magnetic ground states of $EuCd_2As_2$ at different Hubbard energy ($U$) values and external pressures. The magnetic ground state of $EuCd_2As_2$ is closely related with the $U$ values as well as the external pressure. When the Hubbard $U$ values are larger than 3 eV, the AFM_in structure is the most stable at ambient pressure. If $U$ is smaller than 3 eV, the magnetic ground state is always of the FM_out structure regardless of the external pressure, thus exposing the crucial role of the on-site Coulomb interaction. In order to match the normalized lattice constants variation, here $U$ is set as 4 eV. With gradual increase of pressure from 2.1 to 19 GPa, the magnetic ground state experiences the AFM_in, FM_in and finally enters into the FM_out structure. This phase presents magnetic space group $P$-3$m$'1 (164.89 BNS setting) with breaking $\mathcal{T}$ and retaining $P$. The high pressure apparently favors the FM_out structure that could realize the Weyl state, which is tested by Weyl Chirality calculation and magnetic quantum chemistry theory (MQCT) [50, 51]. As is illustrated in Fig. 4(l), linear band crossing points could be well recognized along the G-A high symmetry line, with the crossing points at the position (0.000, 0.000, ±0.0383/Å) and (0.000, 0.000, ±0.0581/Å) which are rather close to $E_F$. The chirality analysis suggests the crossing points are WPs, thus the Weyl state could be regarded as an ideal Weyl state protected by $P$. At the same time, these results are also verified by calculations based on MTQC, which support enforced semimetal along G-A high symmetry line. As shown in Fig. 4(l), the compatibility relations at the three highest valance bands support these two symmetry-protected linear crossing points due to the exchange order of irreducible co-representation at high symmetry points G and A. Apart from these points, there might be a symmetry protected or accidental crossing point with energy position away from $E_F$, which is less concerned. Furthermore, the modified Becke-Johnson method [47] was performed to calculate the Weyl Chirality, which gives the same result as that derived from the DFT+U method.

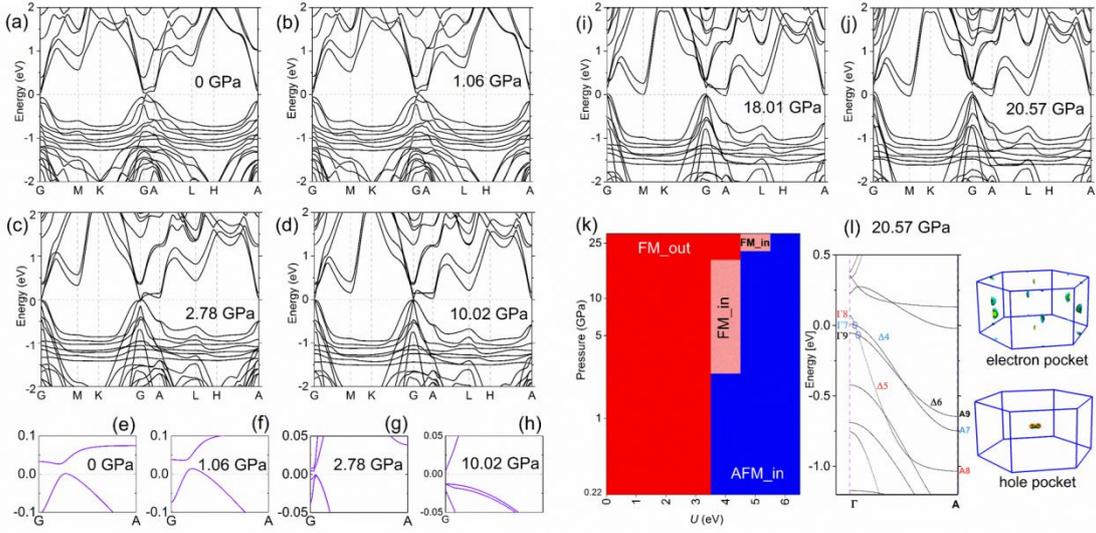

**Fig. 4.** Band structures of the AFM_in magnetic state at (a) ambient pressure and (b) 1.06 GPa, respectively. The electronic band structures of the FM_in magnetic state at (c) 2.78 GPa, (d) 10.02 GPa and (i) 18.01 GPa, and that for the FM_out structure at (j) 20.57 GPa. The band structures for those at low pressure are enlarged in (e) - (h), to clearly show the variation of the band gap. (l) shows the band structure and Fermi surface of $EuCd_2As_2$ at 20.57 GPa. A single pair of WPS could be recognized along the G-A high symmetry line of the Brillouin zone.

## SUMMARY

To summarize, the application of pressure is capable of driving the layered topological semimetal $EuCd_2As_2$ into sequential magnetic states with increasing the pressure, including the AFM_in, FM_in and FM_out structures, neither alters the $CaAl_2Si_2$-type crystal structure nor changes the $Eu^{2+}$ valence state. The analysis of the crystal structure under pressure indicates that the evolutions of the axial ratio (*c/a*) and Eu-layer distance play important roles in driving the magnetic phase transition. The magnetic phase transitions are supported by the first principles calculations. Coupled with the pressure-induced magnetic phase transition, interesting butterfly-shaped MR and anomalous Hall effect were observed. Interestingly, in the FM_out structure, the calculations unveil only one pair of Weyl points near the Fermi level, which could be regarded as an ideal Weyl state. This work demonstrates that external pressure can also realize the ideal Weyl state that ever produced by the application of external magnetic field on $EuCd_2As_2$. The result would be instructive for the discovery of more novel exotic topological properties in magnetic topological phases by using high pressure.

## Acknowledgements

The authors acknowledge the support by the National Natural Science Foundation of China (Grant Nos. 92065201, 11874264, 12134018, 11921004) and the Shanghai Science and Technology Innovation Action Plan (Grant No. 21JC1402000). Y.F.G. acknowledges the start-up grant of ShanghaiTech University and the Program for Professor of Special Appointment (Shanghai Eastern Scholar). Z.C.Z. is supported by Science Center of the National Science Foundation of China (52088101). J.G.C. is supported by the National Key R&D Program of China (2018YFA0305700), Beijing Natural Science Foundation (Z190008), the National Natural Science Foundation of China (Grant Nos. 12025408, 11921004). The high-pressure experiments with cubic anvil cell were performed at the Synergic Extreme Condition User Facility (SECUF). Part of the experiments was performed at the BL15U1 beamline of Shanghai Synchrotron Radiation Facility (SSRF) in China, and the 16 ID-B, Advanced Photon Source (APS) of Argonne National Laboratory (ANL). The synchrotron X-ray absorption spectrum was performed at 20 BM-B of APS at ANL. We thank Steve. M. Heald *et al*. for help with experiment. The authors also thank the support from the Analytical Instrumentation Center (Grant No. SPST-AIC10112914), SPST, ShanghaiTech University.

# Supporting Information

**Pressure-induced ideal Weyl semimetal state in the layered antiferromagnet EuCd$_2$As$_2$**

Zhenhai Yu[1,2,†], Xuejiao Chen[3,†], Wei Xia[1,4,†], Ningning Wang[5], Xiaodong Lv[3], Xiaolei Liu[1], Hao Su[1], Zhongyang Li[1], Desheng Wu[5], Wei Wu[5], Ziyi Liu[5,6], Jinggeng Zhao[6], Mingtao Li[2], Shujia Li[2], Xin Li[2], Zhaohui Dong[7], Chunyin Zhou[7], Lili Zhang[7], Xia Wang[8], Na Yu[8], Zhiqiang Zou[8], Jianlin Luo[5], Jinguang Cheng[5], Lin Wang[2,9,*], Zhicheng Zhong[3,*], and Yanfeng Guo[1,4,*]

[1]School of Physical Science and Technology, ShanghaiTech University, Shanghai, 201210, China

[2]Center for High Pressure Science and Technology Advanced Research, Shanghai, 201203, China

[3]CAS Key Laboratory of Magnetic Materials and Devices & Zhejiang Province Key Laboratory of Magnetic Materials and Application Technology, Ningbo Institute of Materials Technology and Engineering, Chinese Academy of Sciences, Ningbo 315201, China

[4]ShanghaiTech Laboratory for Topological Physics, ShanghaiTech University, Shanghai 201210, China

[5]Beijing National Lab for Condensed Matter Physics, Institute of Physics, Chinese Academy of Sciences, Beijing 100190, China

[6]School of Physics, Harbin Institute of Technology, Harbin 150080, China

[7]Shanghai Synchrotron Radiation Facility, Shanghai Advanced Research Institute, Chinese Academy of Sciences, Shanghai 201204, China

[8]Analytical Instrumentation Center, School of Physical Science and Technology, ShanghaiTech University, Shanghai 201210, China

[9]Center for High Pressure Science (CHiPS), State Key Laboratory of Metastable

Materials Science and Technology, Yanshan University, Qinhuangdao 066004, China

[†]The authors contributed equally to this work.

*Corresponding authors:

wanglin@ysu.edu.cn (LW),

zhong@nimte.ac.cn (ZCZ),

guoyf@shanghaitech.edu.cn (YFG)

## 1. Single crystal growth and basic characterizations

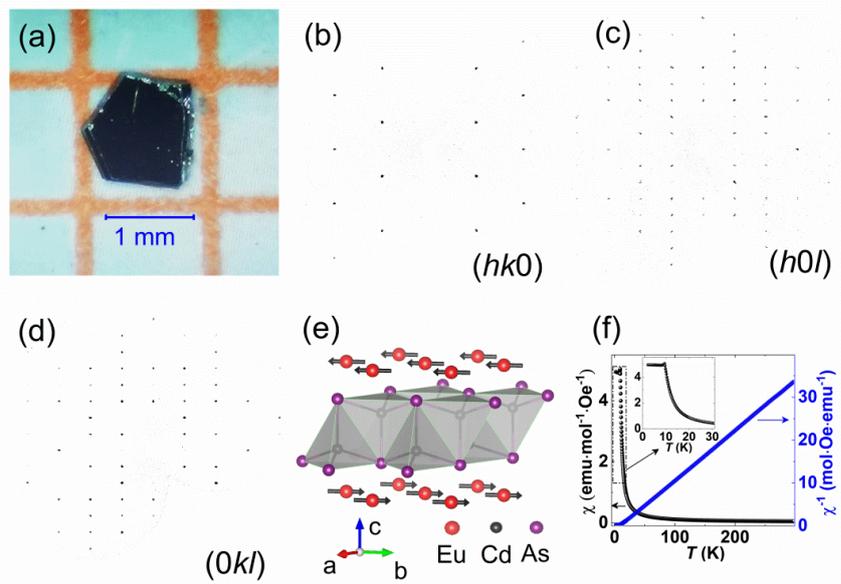

Fig. S1. (a) The picture of $EuCd_2As_2$ single crystal. (b)-(d) Single crystal X-ray diffraction patterns in the reciprocal space along ($hk0$), ($h0l$), and ($0kl$) directions, respectively. (e) The schematic crystal structure of $EuCd_2As_2$ with $CdAs_4$ tetrahedron and the arrows on the Eu ions denote the spin direction. (f) Temperature dependence of the magnetic susceptibility of $EuCd_2As_2$ measured in an external field of 1000 Oe. The blue line represents the reciprocal magnetic susceptibility. The inset shows the enlarged view of the χ data below 30 K.

The $EuCd_2As_2$ single crystals were grown by using a Sn flux method with the procedure similar as that reported reference [10]. The picture for plate-like black crystal with shining surface and a typical size of $1 \times 1 \times 0.2$ mm$^3$ is shown in Fig.

S1(a). The crystal quality was examined on a Bruker D8 single crystal X-ray diffractometer with $\lambda$ = 0.71073 Å at room temperature. The obtained lattice parameters are $a$ = 4.4423 Å and $c$ = 7.333 Å, which are consistent with those reported reference [10]. The clean reciprocal diffraction patterns shown in Figs. S1(b)-1(d) without other impurity spots indicate the high quality of our single crystals. The schematic crystal and magnetic structures of $EuCd_2As_2$ was shown in Fig. S1(e). Temperature dependence of the magnetic susceptibility of $EuCd_2As_2$ measured in an external field of 1000 Oe was shown in Fig. S1(f).

**2. Magnetic properties measurement under ambient and high pressure**

The direct current (dc) magnetization (*M*) for $EuCd_2As_2$ was measured in a Quantum Design magnetic property measurement system (MPMS). The Néel temperature $T_N$ of ~ 9.5 K was derived from the *M-T* data, as shown in Fig. S1(f), consistent with the previously reported value [10]. The magnetizations measurements under high pressure were performed using a miniature BeCu piston-cylinder cell measured on a commercial magnetic property measurement system from Quantum Design. The $EuCd_2As_2$ single crystals together with a piece of Sn were loaded into a Teflon capsule filled with Daphne 7373 as the pressure transmitting medium (PTM). Pressure at low temperatures was determined from the shift of the superconducting transition temperature of elemental Sn. All of the magnetization curves are measured in the zero-field-cooling mode.

**3. High-pressure low-temperature electrical transport measurement**

A symmetric diamond anvil cell (DAC) with an anvil culet of 400 μm in diameter was utilized to generate high pressure. A four-probe method by using Pt as the electrodes was used for in-situ electrical resistance measurements under pressure. The Hall effect measurement under pressure was only performed at 5 K and 100 K. The manufacturing process of the microcircuit was similar to that described in our previous work [37]. The PTM was not used for the resistance measurement. A ruby ball was used as the pressure calibrator [38].

## 4. High pressure angle dispersive X-ray diffraction (AD-XRD) measurements

The $EuCd_2As_2$ crystals were ground in a mortar in order to obtain fine powder sample that is used for high pressure AD-XRD and X-ray absorption near-edge spectroscopy (XANES) measurements. The high pressure AD-XRD measurement was carried out using a symmetric DAC. The AD-XRD patterns were taken with a MarCCD detector using synchrotron radiation beams monochromatized to a wavelength of 0.6199 Å at the beamline BL15U1 of Shanghai Synchrotron Radiation Facility (SSRF). Run #1 was performed at BL15U1 with silicone oil as PTM up to 27.1 GPa. Another independent Run #2 AD-XRD measurement up to 50.0 GPa was carried out at the beamline of 16-ID-B ($\lambda = 0.4066$ Å) at the Advanced Photon Source (APS) of Argonne National Laboratory. The two-dimension image plate patterns were converted into the one-dimension intensity versus degree data by using the Fit2D software package [39]. The experimental pressures were determined by the pressure-induced fluorescence shift of ruby [38]. The XRD patterns were analyzed with Rietveld refinement using the GSAS program package [40] with a user interface EXPGUI [41].

In Figs. S2(a)-S2(b), we present the AD-XRD patterns collected in the two independent experiments. Seen from Fig. S2(a), we found a shift of the Bragg peaks to higher angles upon the increase of pressure, implying shrinkage of the unit cell, while all the XRD patterns could well indexed on the basis of the $CaAl_2Si_2$-type structure, safely excluding any structural phase transition within the pressure range. The typical Rietveld refinement results were shown in Figs. S3(a)-S3(b).

The results of Run #2 AD-XRD experiment are presented in Fig. S2(b). To clearly see the evolution of the Bragg peaks, they were marked below the XRD patterns for several selected pressure. As seen from Fig. S2(b), the XRD patterns below 30.6 GPa can be still indexed by the parent structure with the space group *P*-3*m*1. However, new diffraction peaks appear when the pressure is above 35.5 GPa, likely indicating a pressure-induced structural phase transition. To clarify that the new peaks are not splitted from the parent phase, the (102) and (003) diffraction peaks of the parent structure were labeled and were found to gradually shift with increasing the pressure. Moreover, it is found that the intensity of the new peaks labeled as HP phase apparently increases with increasing the pressure. However, the major diffraction peaks still belong to the parent $P$-$3m$1 structure with the pressure even up to 50.0 GPa. The mixed $P$-$3m$1 and HP phases in the AD-XRD patterns bring great difficulty in performing Rietveld analysis. The parent phase was recovered when the pressure is released to ambient pressure, which is labeled as D0 in top of Fig. S2(b).

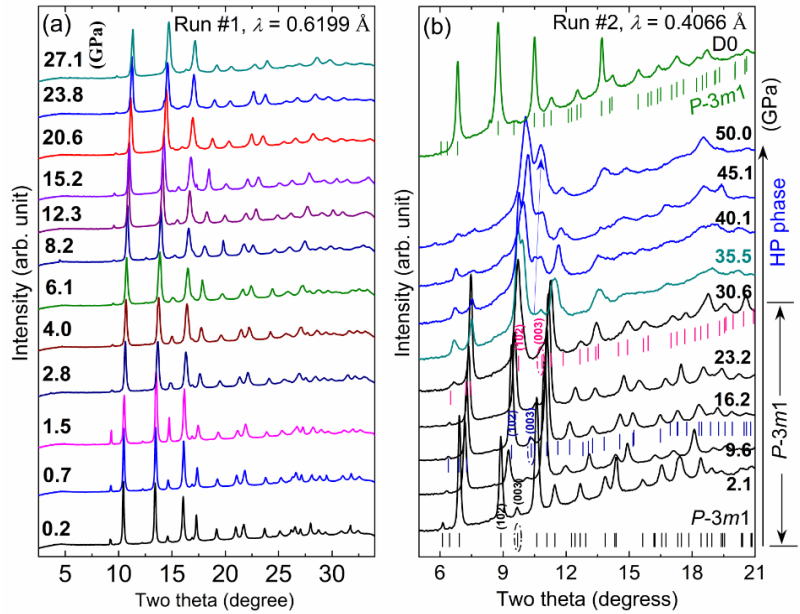

Fig. S2. AD-XRD patterns of EuCd$_2$As$_2$ under various pressures. (a) Run #1 measurement up to 27.1 GPa performed at BL15U1 of SSRF. (b) Run #2 measurement up to 50.0 GPa carried out at 16 ID-B of APS.

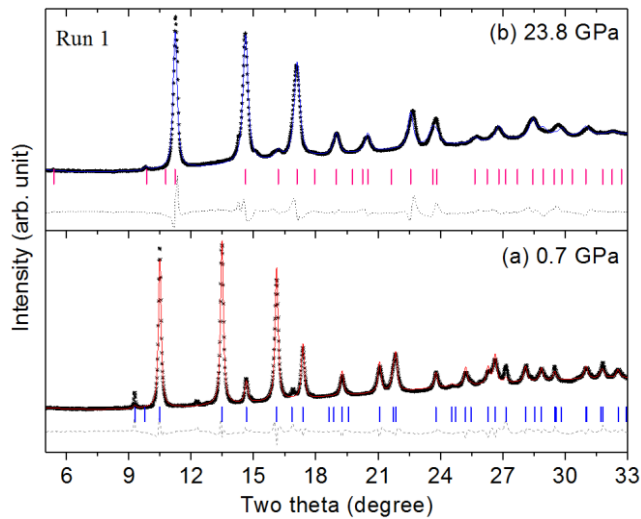

Fig. S3. Typical Rietveld refinement results of the AD-XRD patterns of EuCd$_2$As$_2$ collected in the Run #1 measurement at (a) 0.7 and (b) 23.8 GPa, respectively. The vertical bars represent the calculated peaks of EuCd$_2$As$_2$. The difference between the observed (crosses) and the fitted (line) patterns is shown with a blue line at the bottom.

## 5. XANES of Eu L$_1$ edge under high pressure

High-pressure XANES experiments at Eu L$_1$–edge were performed at beamline 20-BM-B at APS of ANl. The XAS measurements were performed in the transmission mode. The incident beam was monochromatized with two Si (111) single crystals. Silicone oil was used as the PTM. The sample pressure was determined using the standard ruby fluorescence technique [38]. To reduce the absorption from the diamond anvils, nanopolycrystal diamond was used.

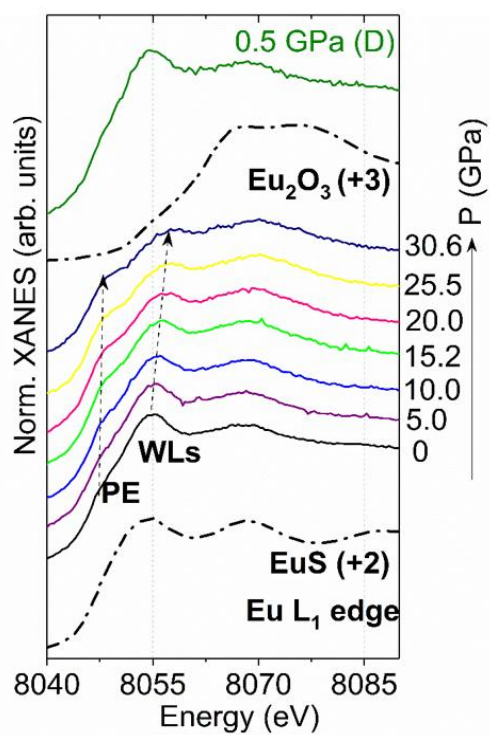

Fig. S4. XANES spectra at ambient temperature for Eu at selected pressures up to 30.6 GPa. The dashed lines are guide to the eyes. The top spectrum corresponds to the XANES collected during decompression from 30.6 GPa. The spectra of EuS (+2) and Eu$_2$O$_3$ (+3) [42] are shown for a comparison.

## 6. DFT calculation

Pressured-induced structural and electronic properties of $EuCd_2As_2$ were simulated with the projector augmented wave method [43] and Perdewe-Burke-Ernzerhof revised for solids (PBEsol) [44] scheme based on Vienna *Ab initio* simulation package (VASP) [45]. The Brillouin zone was sampled with $12 \times 12 \times 4$ (AFM) and $12 \times 12 \times 8$ (FM) Monkhorst-Pack k-mesh and kinetic energy cutoff was set to 500 eV. Experimental lattice parameters were used for calculating the magnetic ground states at the different pressures with setting the on-site Coulomb interactions of Eu 4*f* electrons from 0 eV to 6 eV. The spin-orbital coupling was included for a self-consistence. Eu 4*f*5*d*, Cd 5*s* and As 4*p* orbitals were projected onto maximally localized Wannier functions based on VASP2WANNIER interface [46]. This projected process was also performed with modified Becke-Johnson method [47]. Corresponding topological properties were calculated with WANNIERTOOLS [48] and the Fermi surface was plotted by FermiSurfer [49]. Magnetic topological quantum chemistry method was also used to calculate the compatibility relations by using MagVasp2trace package [50, 51].